\documentclass[twocolumn,aps,prl,showpacs]{revtex4-1}
\usepackage{epsfig}
\usepackage{graphics}
\usepackage{amssymb}
\usepackage{amsmath}

\newcommand{\nl}{\nonumber\\}

\newcommand{\ha}{\hat{a}}

\newcommand{\hb}{\hat{b}}

\begin{document}

\title{Antibunched Emission of Photon-Pairs via Quantum Zeno Blockade}

\author{Yu-Ping Huang}
\affiliation{Center for Photonic Communication and Computing, EECS Department\\
Northwestern University, 2145 Sheridan Road, Evanston, IL 60208-3118}
\author{Prem Kumar}
\affiliation{Center for Photonic Communication and Computing, EECS Department\\
Northwestern University, 2145 Sheridan Road, Evanston, IL 60208-3118}

\begin{abstract}
We propose a new methodology, namely ``quantum Zeno blockade,'' for managing light scattering at a few-photon level in general nonlinear-optical media, such as crystals, fibers, silicon microrings, and atomic vapors. Using this tool, antibunched emission of photon pairs can be achieved, leading to potent quantum-optics applications such as deterministic entanglement generation without the need for heralding. In a practical implementation using an on-chip toroidal microcavity immersed in rubidium vapor, we estimate that high-fidelity entangled photons can be produced on-demand at MHz rates or higher, corresponding to an improvement of $\gtrsim10^7$ times from the state-of-the-art.
\end{abstract}
\maketitle

Generation of quantum entanglement is an interdisciplinary, long-lasting effort, triggered more than fifty years ago by Bell's quantum non-locality argument \cite{BellInequality64} in response to the hidden-variable theory of Einstein, Podolsky, and Rosen \cite{EPR35}. Motivated by the fundamental tests of quantum uncertainty in earlier days, the quest for efficient sources of entanglement nowadays has been fueled by a variety of potent applications that are otherwise unrealizable by classical means (see Ref.~\cite{Horodecki09} for a review). For most of these applications, entanglement embodied in pairs of photons has been recognized as an ideal resource owing to its robustness against decoherence, the convenience of its manipulation with linear-optical components, as well as the ease of distribution over long distances at the speed of light. Thus far, entangled photon pairs have mostly been generated probabilistically via post-selection \cite{note1}, where the quantum-entanglement features are established only after selecting favorable measurement outcomes. While such photon pairs are useful for some proof-of-principle demonstrations of quantum effects, practical applications beyond a few-qubit level will require on-demand sources of entangled photons.

The obstacle to deterministic generation of entangled photons in nonlinear-optical media arises fundamentally from the stochastic nature of the photon-pair emission process, because of the inherent quantum randomness in how many photon pairs will be created in a given time interval \cite{StatisticPDC00}. To overcome this randomness, existing methods have relied on ``heralding'' schemes in which auxiliary photons are detected in order to project a multi-photon-pair state onto an entangled single-pair state \cite{KokBra00}. In these schemes, however, a four-fold coincidence measurement \cite{HEPAN10,HeZei10} or a two-fold coincidence measurement after nonlinear-optical mixing must be adopted \cite{SFGQKD11}. Because such operations are extremely inefficient, the production rate of entangled photons is fundamentally restricted to the \emph{sub-Hertz} range.

In this Letter, we propose and demonstrate via simulation a new methodology for managing light scattering in general nonlinear media, which allows us to directly overcome (i.e., without the use of heralding) the stochastic nature of the photon-pair emission process. The idea is to employ novel ``quantum Zeno blockade'' (QZB), which suppresses the creation of multiple photon pairs in a single spatiotemporal mode through the quantum Zeno effect \cite{Zeno77}, while the creation of a single pair is allowed. It is achieved by coupling the photon-pair system to a dissipative reservoir in a way that the coupling is efficient only when more than one pair of photons are present. When the coupling is sufficiently strong, the creation of multiple photon pairs is then blocked (suppressed) through the quantum Zeno effect \cite{HuaMoo08}. As a result, the photon pairs are created in a pair-wise ``antibunching'' manner similar to that of antibunched emission of single photons by a single atom \cite{ScuZub97}. Such can lead to deterministic generation of entangled photons at \emph{MHz} rates or higher by using existing technology, an example of which will be shown later in this Letter. We note that while QZB relies on a strong coupling between multiple pairs of photons and a reservoir, but when it is in effect, ideally no energy dissipation or quantum-state decoherence will actually occur as the creation of multiple pairs will be inhibited.

We consider implementing QZB via two-photon absorption (TPA). Other approaches, such as that via stimulated four-wave mixing, are also possible. TPA is a nonlinear-optical phenomenon in which two overlapping photons are simultaneously absorbed, while the absorption of any one of them alone is inhibited, i.e., occurs with a much lower efficiency. TPA has been studied for decades and successfully demonstrated in a variety of physical systems, including ion-doped crystals \cite{EuDopeTOA61}, atomic vapors \cite{TPACs62}, semiconductors \cite{VanIbrAbs02}, and molecules \cite{TPAMelocule00}. For generating antibunched photon-pairs, we employ the degenerate TPA process wherein two photons of the same wavelength are absorbed simultaneously. When the TPA-induced QZB is in effect, the creation of a single photon pair prevents additional pairs from being created in the same spatiotemporal mode via the quantum Zeno effect. In this respect, the proposed QZB is analogous to the dipole blockade in Ryberg-atom systems \cite{Dipoleblockade01} or the photon blockade in atom-cavity systems \cite{PhotonBlockade05}. There is, however, a distinct difference. Both the dipole blockade and photon blockade result from coherent energy-level shifting created by ``real'' potentials, which in these two cases are caused by dipole-dipole interaction and vacuum Rabi splitting, respectively. In contrast, QZB is realized through energy-level broadening produced by an ``imaginary'' pseudo potential caused by an incoherent, dissipative TPA process (see Ref.~\cite{HuaAltKum10-2} for a detailed comparison of the two effects).

\begin{figure}
\centering \includegraphics[width=5.5cm]{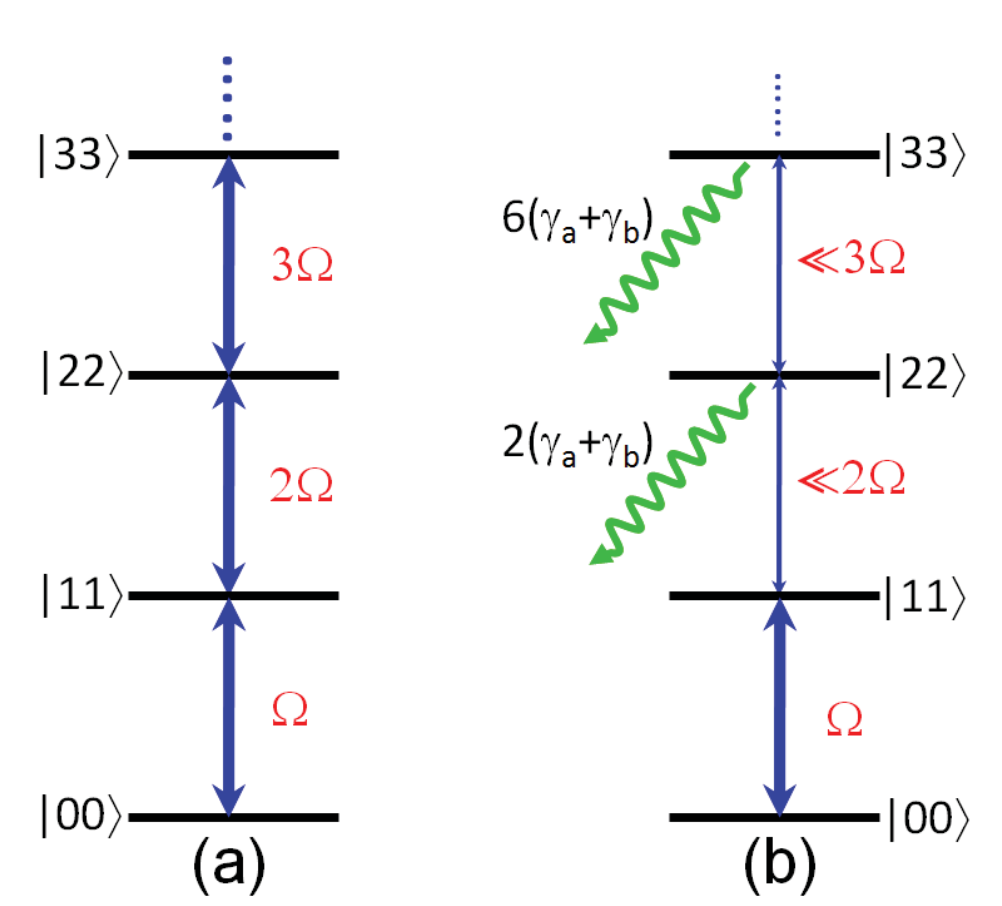}
\caption{(Color online) Level-transition diagrams for photon-pair generation in the absence of TPA (a) and when strong TPA-induced QZB is present (b). \label{fig1}}
\end{figure}

To study the QZB-caused antibunched emission of photon pairs, we consider a model whose level-transition diagram is drawn in Fig.~\ref{fig1}. Such a model can be generally applied to a variety of nonlinear optical processes, such as spontaneous parametric downconversion (SPDC), spontaneous four-wave mixing (SFWM), and resonant two-photon superradiance. In this model, photon pairs are generated through a phase-matched wave-mixing process governed by the following interaction Hamiltonian:
\begin{equation}
    \hat{H}_\mathrm{pair}=\hbar \Omega \ha^\dag\hb^\dag+\mathrm{H.c.},
\end{equation}
where $\Omega$ is a real constant measuring the pair-generation efficiency, and $\ha^\dag$ and $\hb^\dag$ are the usual creation operators for generating photons in the signal and idler modes, respectively. The TPA process is treated via a master-equation approach \cite{ScuZub97}, resulting in the following equation of motion for the system density matrix
\begin{eqnarray}
\label{meq}
    \frac{d\rho}{d\xi}&=&\frac{i}{\hbar}[\rho,\hat{H}_\mathrm{pair}] +\frac{\gamma_a}{2}(2\ha^2\rho\ha^{\dag 2}-\ha^{\dag 2}\ha^2\rho-\rho\ha^{\dag 2}\ha^2)
   \nl
    & & +\frac{\gamma_b}{2}(2\hb^2\rho\hb^{\dag 2}-\hb^{\dag 2}\hb^2\rho-\rho\hb^{\dag 2}\hb^2),
\end{eqnarray}
where $\xi$ is a moving-frame coordinate, $\gamma_a$ ($\gamma_b$) is the TPA coefficient for the signal (idler) photons, and the linear loss of the signal and idler photons is assumed to be negligible compared to their nonlinear loss via TPA.

The system dynamics governed by Eq.~(\ref{meq}) is visualized in Fig.~\ref{fig1}, where the Dirac notation $|00\rangle,|11\rangle, |22\rangle \cdots$ labels the number states containing zero, one, two, $\cdots$ pairs of photons, respectively, in the signal-idler modes. Without TPA ($\gamma_a=\gamma_b=0$), as shown in Fig.~\ref{fig1}(a), ``ladder''-like energy states are successively excited. Starting with the vacuum state $|00\rangle$, an infinite sequence of states containing one, two, $\cdots$ pairs of photons can be populated. With TPA, in contrast, the higher-order processes in the ladder transitions involving $|11\rangle\leftrightarrow|22\rangle$, $|22\rangle\leftrightarrow|33\rangle$, etc., are suppressed, as shown in Fig.~\ref{fig1}(b). When TPA is sufficiently strong, the $|00\rangle$ and $|11\rangle$ states form an isolated Hilbert sub-space, and the transition dynamics corresponds to a Rabi oscillation between these two states. For $\gamma\equiv\gamma_a+\gamma_b\gg \Omega, 1/L$ ($L$ is the effective interaction length for photon-pair generation), the system dynamics (\ref{meq}) can be solved approximately via adiabatic elimination, giving ($P_n$ is the probability to create $n$ pairs of photons)
\begin{eqnarray}
\label{p1p2}
    P_1 \simeq \sin^2(\Omega L), ~~
    P_2 \simeq (2\Omega/\gamma)^2 P_1\ll P_1^2, ~~\cdots,
\end{eqnarray}
which exhibits the characteristic of pair-wise antibunching. Photon pairs possessing such statistical properties can be used in a variety of quantum-information applications that are operated on demand, such as deterministic entanglement swapping without post selection and heralded generation of entangled photon pairs using only linear-optical instruments. Even for those applications not requiring event-ready entangled photons \cite{Rev-Qua-cryp}, such photon pairs can significantly improve the rate at which such applications can be operated by substantially reducing the background noise arising from multi-pair emission \cite{HuaAltKum11}.

Particularly, in Eq.~(\ref{p1p2}), when $\Omega L=\pi/2$ the probability to create a pair of photons is $P_1\simeq1$. The probability to create multiple pairs, on the other hand, is about $4\Omega^2/\gamma^2\ll 1$. Thus, on-demand entangled photon pairs are created directly with high fidelity, without the need for any post pair-generation procedure such as heralding. The pair-production rate can therefore be very high, e.g., tens or hundreds of MHz, limited only by the bandwidth of the photon pairs. Such rates would correspond to an improvement by more than $10^7$ times over those achievable via the existing methods \cite{HEPAN10,HeZei10}.

\begin{figure}
\centering \includegraphics[width=7cm]{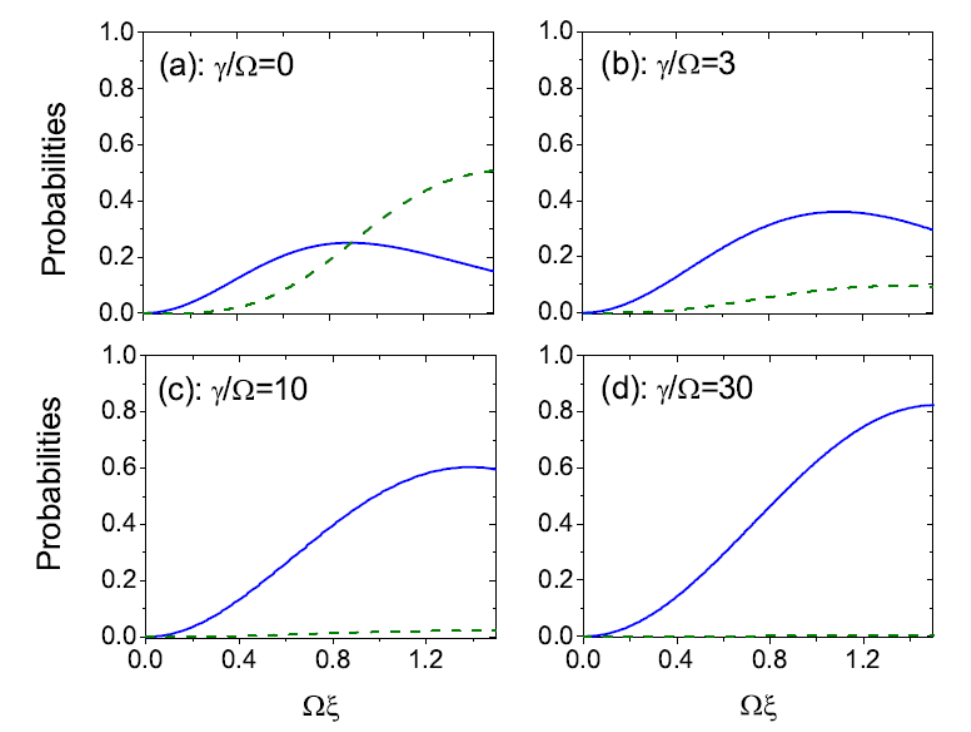}
\caption{(Color online) Probabilities to create a single pair of photons $P_1$ (solid) and multiple pairs $P_{n>1}$ (dashed), plotted as function of $\Omega\xi$ for several TPA absorption strengths $\gamma$. \label{fig2}}
\end{figure}
To ratify these analyses, we perform numerical simulations for the case with $\gamma_b=0$, i.e., only the signal photons are subjected to TPA but not the idler photons, as such is expected to be easier to achieve in experimental implementations. The simulation results are plotted in Fig.~\ref{fig2}, where in (a)--(d) $\gamma/\Omega$ equals $0,3,10,30$, respectively, corresponding to an increasingly stronger QZB effect. In Fig.~\ref{fig2}(a) without TPA, the probability to create multiple photon pairs $P_{n>1}\equiv\sum_{n\ge 2} P_n$ is about $P_1^2$ in the weak-pump regime when $\Omega \xi \ll 1$, as expected. When the pump power is increased, $P_{n>1}$ increases much faster than $P_1$. The two probabilities then intersect at $\Omega\xi=0.9$, for which $P_1=0.25$. When TPA is present, however, the pair-generation dynamics is modified due to the QZB effect. For moderate $\gamma=3\Omega$, the probability to create multiple photon pairs is already suppressed, as shown in Fig.~\ref{fig2}(b). For stronger TPA, the probability to generate multi-pairs is substantially reduced. With $\gamma/\Omega=10$, as shown in Fig.~\ref{fig2}(c), $P_1=0.6$ and $P_{n>1}=0.026$ are obtained for $\Omega\xi=1.4$, corresponding to suppression of multi-pairs by 34 times below a typical non-antibunched result \cite{StatisticPDC00}. With $\gamma/\Omega=30$, as shown in Fig.~\ref{fig2}(d), $P_1=0.83$ and $P_{n>1}=0.0036$ are achieved for $\Omega\xi\approx \pi/2$, establishing an ultra-strong pair-wise antibunching effect. Note that by introducing TPA for the idler photons as well (i.e., $\gamma_b>0$), the pair-wise antibunching effect can be significantly enhanced. Finally, comparing Figs.~\ref{fig2}(a)--(d), we emphasize that for a stronger TPA channel, the peak production rate of single photon-pairs that can be achieved is increased. This behavior reflects the fact that enhanced QZB provides a better isolation for the Rabi-oscillation dynamics in nonlinear media.

\begin{figure}
\centering \includegraphics[width=7.0cm]{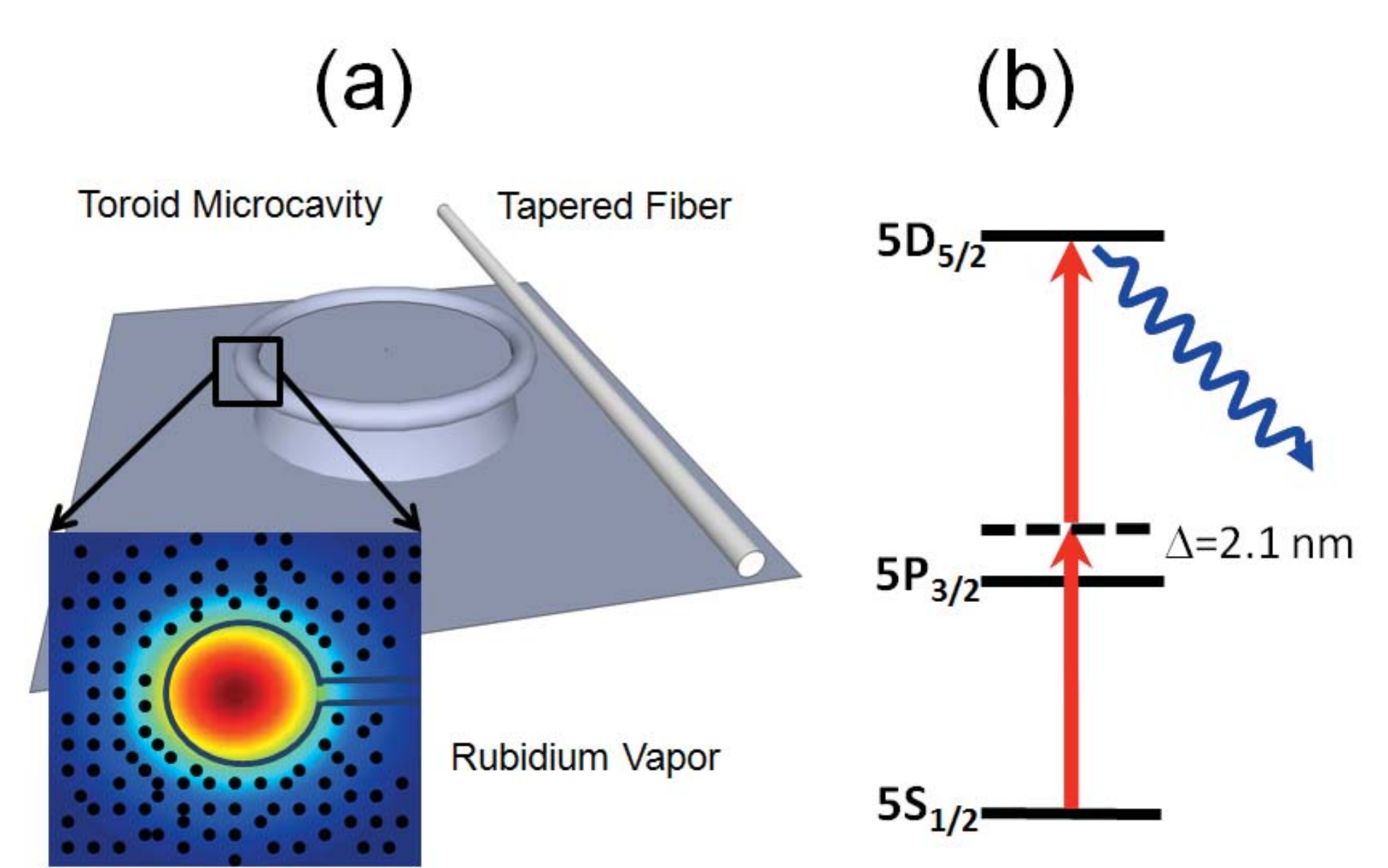}
\caption{(Color online) (a) A schematic of the photon-pair source made of a microcavity immersed in a Rb-vapor cell. (b) The level-scheme in Rb atoms for the degenerate TPA process. \label{fig4}}
\end{figure}

We now present a practical implementation of the QZB-caused antibunched emission of photon pairs using an on-chip toroidal microcavity, whose fabrication techniques and applications in nonlinear optics have been well developed \cite{HighQMicroring03,Microring05}. The device schematic is shown in Fig.~\ref{fig4}(a). Basically, the cavity consists of a Kerr-nonlinearity microring fabricated on top of a silicon pedestal, with light waves guided along the microring's periphery. The microring is coupled to a tapered fiber via an evanescent interface, with a coupling $Q$-factor arranged to be much less than the cavity's intrinsic quality factor $Q_i$ so as to avoid loss in the cavity. Thus far, cavities of this kind have been fabricated with ring diameters as low as $\sim50~\mu$m, and $Q_i$ well above $10^8$. For photon-pair generation, the microcavity geometry is arranged to achieve both triple-resonance and phase-matching for the pump, the signal, and the idler light waves. Such a technique has also been demonstrated in experiment \cite{KipSpiVah04,FerRazDuc08}. For this Letter, we consider the signal photon to be at $778$ nm, while the pump and the idler are well detuned from Rb transition lines.

To achieve QZB, the microcavity is immersed in a Rb-vapor cell. TPA for the signal photons is thus achieved via evanescent coupling to Rb atoms close to the microring surface, as illustrated in the inset of Fig.~\ref{fig4}(a). The atomic energy-level scheme accounting for this TPA process is sketched in Fig.~\ref{fig4}(b), where excitations from $5S_{1/2}$ to $5P_{3/2}$ and from $5P_{3/2}$ to $5D_{5/2}$ are successively driven by two signal photons. Because of a relatively small ($2.1$ nm) intermediate-level detuning, a large TPA cross-section can be obtained. Thus far, strong TPA in Rb vapors has been observed in the system of hollow-core fibers \cite{TPAHollow11} and tapered fibers \cite{TPATapered10}. For a typical toroidal microcavity, it has been shown that a large TPA coefficient on the order of gigahertz is obtainable using a Rb-atom density of $\sim 10^{14} /\mathrm{cm}^3$ \cite{JacFra09}. Furthermore, it has been predicted that the TPA cross-section can be significantly increased by using the electromagnetically-induced transparency (EIT) effect \cite{TPAEIT}.

\begin{figure}
\centering \includegraphics[width=7.0cm]{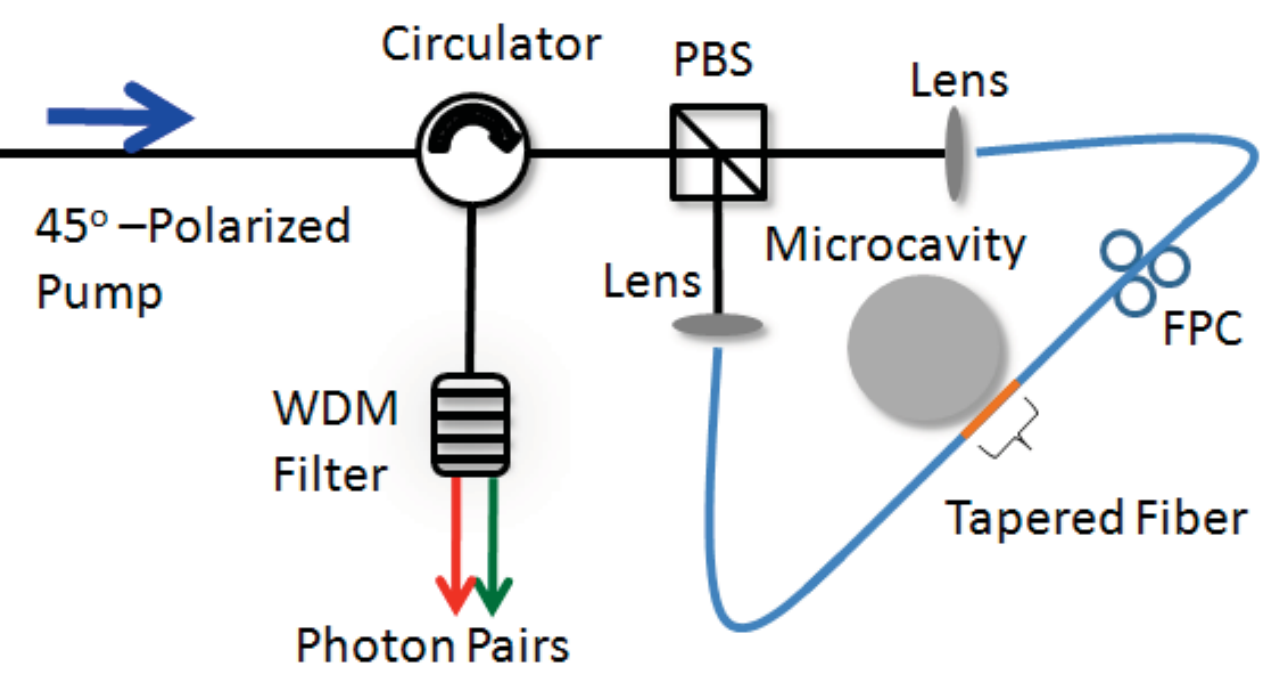}
\caption{(Color online) A schematic setup for deterministic generation of entangled photon pairs using a microcavity evanescently coupled with a Rb vapor. PBS: polarizing beamsplitter; FPC: fiber-polarization controller. \label{fig5}}
\end{figure}

With the above pair-wise antibunched-emission system, entangled photons can be deterministically created adopting, for example, either a counter-propagating (CP) scheme \cite{CPS04,CPSreview08}, a quantum-splitter scheme \cite{Medic10}, or a time-bin scheme \cite{Timebin89}. As an example, a CP scheme for the microring system is schematically depicted in Fig.~\ref{fig5}(a). Briefly, a $45^o$-polarized pump pulse is passed through a polarization beam splitter (PBS) and split equally into horizontal and vertical components. The two components are then propagated along clockwise ($cw$) and anti-clockwise ($acw$) directions, respectively, in a fiber loop. The loop contains a fiber-polarization controller (FPC), which flips the incident polarization from horizontal to vertical and vice-versa. It is also coupled with the microring cavity via a piece of tapered fiber that provides an evanescent coupling interface. By adjusting the fiber-path length and tuning the FPC, the $cw$ and $acw$ pump components are arranged to arrive simultaneously at the cavity with the same polarization. In the cavity, they create equal probability amplitudes for photon-pair emission in the $cw$ and $acw$ propagating modes, respectively, via the Kerr nonlinearity in the microring. The probability to create simultaneously two photon pairs in copropagating or counterpropagating modes, however, are suppressed due to the TPA-induced QZB. The created photon pairs are then coupled out to the fiber through the evanescent interface. The polarization of the $acw$-propagating photons will then be flipped by the FPC, so that when arriving at the PBS, the $cw$ and $acw$ photon pairs are combined to form a single beam in a polarization-entangled state of  $\frac{1}{\sqrt{2}}(|11\rangle_H+|11\rangle_V)$ (up to a controllable relative phase between the two polarizations). The entangled photon pairs are then collected by passing the PBS output through an optical circulator followed by a wavelength-division-multiplexing (WDM) filter.

For this system, deterministic creation of entangle photon pairs can be achieved for $\sqrt{2}\Omega \tau\approx\pi/2$ and $\gamma\gg \Omega$, where $\tau$ is the effective interaction time for the pair-generation inside the cavity. For realistic $\gamma=2$ GHz that is obtainable with a Rb-vapor density of $\sim 10^{14}/\mathrm{cm}^3$ \cite{JacFra09} (or lower if the EIT-enhancement effect is employed \cite{TPAEIT}), $\Omega=0.1$ GHz obtainable with an appropriate pump power, and $\tau=10$ ns achieved by adjusting the microring-fiber coupling, the probability to create a single pair of entangled photons $P_1$ is $0.74$. The probability to create double pairs $P_2$, on the other hand, is only 0.014, exhibiting a strong pair-wise antibunching effect. For an enhanced QZB effect, $P_1$ can be quickly increased to be near unity while $P_2$ is further suppressed. As an example, for $\gamma=10$ GHz, $\Omega=0.1$ GHz and $\tau=11$ ns, $P_1=0.94$ and $P_2=0.0005$ are obtained. In this case, high-fidelity entangled photon pairs are created deterministically without the need for any post pair-generation procedure such as heralding. Note that the actual pair-production rate obtained in practice could be lower due to photon losses arising from, for example, intrinsic cavity loss, fiber-cavity coupling loss, and single-photon scattering by the Rb vapor, the detailed effects of which will be presented elsewhere.

In summary, we have proposed a new methodology, namely quantum Zeno blockade, for overcoming the stochasticity in light-scattering in nonlinear media. Using this tool, we have shown that antibunched photon-pairs in correlated or entangled states can be created deterministically at {\it MHz} rates or higher in a practical microring-cavity system. Our results reveal an avenue to unprecedented phenomena and applications in modern quantum optics, including deterministic (non-post-selected) entanglement swapping performed using only linear-optical instruments, ultra-bright single-photon sources via heralding, and quantum-key distribution with a fresh-key-generation rate substantially higher than the state-of-the-art. We note that the microring-cavity system described in this Letter can also be used for low-loss high-fidelity all-optical logic in both classical and quantum domains.

This research was supported in part by the Defense
Advanced Research Projects Agency (DARPA) under
the Zeno-based Opto-Electronics (ZOE) program (Grant
No. W31P4Q-09-1-0014) and by the United States Air
Force Office of Scientific Research (USAFOSR) (Grant
No. FA9550-09-1-0593).

%

\end{document}